# PLAY OR SCIENCE? A STUDY OF LEARNING AND FRAMING IN CROWDSCIENCE

Andreas Lieberoth, Mads Kock Pedersen, Jacob Friis Sherson
andreas@psy.au.dk, madskock@gmail.com, sherson@phys.au.dk
Aarhus University, Denmark
Centre for Community Driven Research (CODER)
Science and Technology
Ny Munkegade 120
DK-8000 Aarhus C

**Abstract**

Crowdscience games may hold unique potentials as learning opportunities compared to games made for fun or education. They are part of an actual science problem solving process: By playing, players help scientists, and thereby interact with real continuous research processes. This mixes the two worlds of play and science in new ways. During usability testing we discovered that users of the crowdscience game Quantum Dreams tended to answer questions in game terms, even when directed explicitly to give science explanations. We then examined these competing frames of understanding through a mixed correlational and grounded theory analysis. This essay presents the core ideas of crowdscience games as learning opportunities, and reports how a group of players used "game", "science" and "conceptual" frames to interpret their experience. Our results suggest that oscillating between the frames instead of sticking to just one led to the largest number of correct science interpretations, as players could participate legitimately and autonomously at multiple levels of understanding.

**Introduction**

When learning games first entered the scene, curriculum content and teaching methods shifted very little. Surface features of gameplay were added, but drills and narrative construction mirrored what was known on paper, blackboards and older media. Brenda Laurel memorably described this as 'chocolate covered broccoli' (2001): The same old stuff with a game design forced around it, such as getting to fire your gun only after completing a math problem in the *Space Invaders* clone *Math Blaster*.

As purposeful play gained momentum, however, the maturing games industry increasingly came to shape play practices outside "just for fun" contexts. The medium was increasingly shaping the message. Or rather, games are no longer seen as delivery mechanisms for content, but as ecologies of participation.

In this essay, we use the little action game *Quantum Dreams* (http://scienceathome.org/games/quantum-dreams/) to present the learning potentials in crowd science games, where participants are actually helping a scientist by playing. We then discuss the challenge of having mixed epistemic frames in the play experience: The immediate game interface on one hand, and the science process on the other. When used in a classroom setting, a third frame, learning and education, is also added.

This conundrum is unpacked through a grounded and correlational analysis of 38 players' interpretations of interface-elements in Quantum Dreams. The fact that many players seemed to place focus on either game or science surfaced during pragmatic perusing of usability test data, and was turned into a more formal analysis for the sake of this essay.

**Gaming for science**

Sawyer and Smith's "serious games typology" from GDC 2008 identified science and research as one of the seven major purposes that games now serve for various audiences including in healthcare, industry and government (Breuer & Bente, 2010; Klopfer et al., 2009; Sawyer & Smith, 2008).

Crowdscience games represent a tipping point, where serious game playing not just supports changes in attitudes or competences in the user, but makes an active difference for researchers trying to solve a problem – from mapping the neurons of the mouse retina, over curating archaeological artifacts, to building the controlling AI for a quantum computer.

**Citizen science is not new**

It could sound like the crowd science movement was a direct manifestation of the transformative power of games envisioned by utopists like Jane McGonigal (McGonigal, 2011). Its roots, however, are to be found much further back – before the internet, and even before science was segregated from leisure and craft. When Charles Darwin wrote his Origin of the Species and Gregor Mendel got curious about genes in his greenhouse, they were just taking part in the societal agenda of their day. Granted, they had time and means not available to the vast majority of rural denizens and the emerging urban populace, but they were not professional scientists contracted by a university or corporation.

These early citizen scientists were motivated by their own curiosity, needs and times, but there are also early examples of regular people being recruited into centralized efforts. Amateur bird lovers and entomologists have, for instance, long helped track the movement of species across the continents. The advent of modern communication technologies enabled this process further, allowing the Smithsonian Institution to recruit local individuals to maintain weather stations and wire in results, creating a real-time meteorological map of the continental United States.

This was viewed as an opportunity to participate and learn as well as a civic duty.

In this sense, the telegraph foreshadowed what would become online crowd science: Some centralized organizer at e.g. a university or NGO creates and advertises an infrastructure that allows ordinary people with a little time on their hands to contribute.

**Cultural psychological motives for crowd science participation**

Understanding why people would want to contribute to science today must be seen in the light of the frames work and leisure. Industrialization institutionalized work, with payment based on exact measures of time and effort, contrary to the past where the largely rural population worked based on immediate seasonal needs. In essence this new "iron cage of capitalism" created a formal, psychological and cultural separation of leisure from work hours (Weber, 1905/2005).

Humans have played in all cultures that we know of (Avedon & Sutton-Smith, 1971; Huizinga, 1959; Suits, 1972), but with the new wage economy, spare earnings could be spent, and new demands for entertainment and dedicated free time was born. This became a theme in worker's rights. In 1888 hundreds of trade unionists thus paraded through Worchester Massachusetts bearing a banner that read *"eight hours for work, eight hours for rest, eight hours for what we will."* Workers wanted opportunities for recreation (Ashby, 2006). Together with the technological possibilities that first gave us dime novels, cheap sheet music and nickel theaters, this can be viewed as a cornerstone in western culture and its entertainment industry that would lead to the rise of cinema, flow-TV and eventually computer games. As gaming progressed from niche market to mobile mass movement, a new age of casual gaming arose. In the new millennium, women over 30 would be the most rapidly expanding consumer group for years on end, and gaming moved from high-investment titles on stationary screens to little pauses in life (Juul, 2010; Software Entertainment Association (ESA), 2013; Wei & Huffaker, 2012). We are experiencing an unparalleled acceptance of play into everyday life – a ludification of culture (Raessens, 2006) and a cognitive surplus which can be put toward informal education and interesting problem solving (Shirky, 2010).

It is in this context that participation in crowd science projects must be understood. While earlier incarnations of citizen science such as the Smithsonian web of weather stations often required some level of expertise and civic sensibility, online technology places the tools needed to contribute at anyone's fingertips, and strives to shape an engaging learning curve from slight interest (Lieberoth, Kock, Marin, Planke, & Sherson, 2014) using the frame and mechanics of game play.

We now see crowd science games in numerous domains, ranging from our own work in fields like psychology (Lieberoth, 2014a) and physics (Sørensen et al., 2015; Lieberoth et al., 2014, Magnussen, Hansen, Planke & Sherson, 2014, Bjælde, Pedersen & Sherson., 2014) to astronomy (Raddick et al., 2010), protein folding (Cooper et al., 2010) and other STEM-subjects, but also spreading to new exiting areas like transcribing historical texts and fieldnotes (Chrons & Sundell, 2011). No matter the domain, players get the chance to take an active part in solving real problems or curating real materials, getting casually acquainted with the area, materials and real cutting edge problems in the process.

**Crowd science games as learning opportunities**

While some crowd science games mainly exist as game interfaces, most of the institutions behind the genre go to some length to inform users about the scientific project they will be contributing to, and even build educational elements into the game architecture.

This is especially important to games where a modicum of skill is needed to really contribute. For instance, our early game *Quantum Moves* required quite a bit of training before users could traverse the difficult levels that represented truly wicked problems in building our quantum computer, compared to how new users can contribute to Galaxy Zoo straight away, even if they may become more speedy and precise with practice (Lieberoth et al., 2014).

As such, crowd science games can be educational in their own right, but we believe that their true educational potential lies as part of *a game based pedagogy* rather than as a stand-alone deployment device for learning practice. There is perhaps a naïve conception in educational game design, that participation alone is enough to engender learning. Time spent on any task will bolster skills and some concepts may transfer near-automatically. However, it is nontrivial to align the activity in a way that allows the player to gain some immediate payoff while creating a sustained and meaningful learning trajectory (Dewey, 1938a; Dreier, 2003; Squire, 2006).

Game experiences with real science allows teachers to solidify teachable moments and weave cognitive hooks into their existing teaching agendas (Avery, 2008; Davis, Horn, & Sherin, 2013; Haug, 2014; Lieberoth & Hansen, 2011)

Having the game awkwardly wrangled onto the content "edutainment"-style is generally considered bad design (Charsky, 2010; Klopfer et al., 2009; Resnick, 2004). We suggest that crowd science games supply an advantage with regard to this challenge, as there is less disjunction between the medium and the science matter – the context and the content are both scientific, and accumulating data demonstrates that this attracts people with just a casual interest in e.g. quantum physics to corresponding games. The crowd science game supplies a genuine opportunity for legitimate peripheral participation (as per Lave & Wenger, 1991) in the scientific process. User engagement may be bolstered through the gameplay itself, or as is often seen via a wider ecology of knowledge of information, interesting quizzes, social milieus, and even opportunities to co-create the game itself.

An analytical approach to these challenges would be to analyze the epistemic frames – games versus science – under which the activity is interpreted by different users, and assess if the two interpretations can coexist in parallel, as supports for one another, or not at all. When we discovered that these levels were clearly dissociable in a set of usability surveys from an educational play session, we decided to investigate further. This is the subject of the remaining parts of this paper.

**Game well played or science done well? A question of framing**

So what defines the play experience of a crowd science game? Viewing crowd science games through the standard motivational frameworks (Huizenga, Admiraal, Akkerman, & Ten Dam, 2009; R. M. Ryan, Rigby, & Przybylski, 2006; Wouters, van Nimwegen, van Oostendorp, & van der Spek, 2013) and player types (Hamari & Tuunanen, 2014) offers some useful design heuristics and measuring tools, but this only seems to paint half the picture. Dropout and conversion rates in *Quantum Moves* resembled most free to play games (i.e. Draganov, 2014; Fields, 2014), but deeper analyses from *Galaxy Zoo* and *The Milky Way Project* revealed that engagement profiles could be sorted into types ranging from briefly hardworking, over moderate, to lasting categories, which cannot simply be boiled down to gameplay (Ponciano, Brasileiro, Simpson, & Smith, 2014).

Indeed, recent research has shown that framing the same activity as either game or work irrespective of the game elements used can have a measurable psychological impact in terms of interest and enjoyment in the short term (Lieberoth, 2014b).

Accordingly, interview studies have shown that while game elements attract new users to citizen science platforms, they are less of a factor in sustained engagement (Iacovides, Jennett, Cornish-Trestrail, & Cox, 2013; Lieberoth et al., 2014)

Framing thus seems to be a central issue: If players view a crowd science game only in comparison with other online games, they will often be disappointed. However, if part of their interest stems from *or* shifts to intrinsic motivation related to taking part in the science project, then play and science frames can merge into a new level of enjoyable experiences. To understand a well played citizen science game, we must thus try to understand not just the raw game play, but also the meta-motivational frame under which the activity unfolds, and how this shapes players' interpretation of the game elements.

**Quantum Dreams: a play experience analysis**

To put the discussion presented above under scrutiny, we examined user experiences in the crowd science game Quantum Dreams. Quantum computers offer immense computational speedup compared to conventional computers by replacing bit, which can be either 0 or 1, with qubits. These can be both 0 and 1 at the same time. Thus, a quantum computer with N qubits can represent 2N different values at the same time, allowing an exponential increase in the computing power for certain tasks (Nielsen & Chuang, 2000). Our approach is to build a quantum computer from ultra-cold atoms in an optical lattice (Weitenberg, Kuhr, Mølmer, & Sherson, 2011). The individual atoms are transported around the lattice by optical tweezers. However, when moved the atoms begin to slosh – similar to coffee in a cup if you are not careful. Computer algorithms are only capable of solving the problem of transporting the atom without sloshing, if given enough time. To investigate whether humans given the right visual tools can form heuristic algorithms to find fast solutions to the complex quantum problem of moving a single atom without sloshing, we built the game Quantum Dreams in the Unity game engine. Quantum Dreams represents a simple 3D game loop based on the more complex levels in our less smooth game Quantum Moves (Sørensen et al., 2015; Lieberoth et al., 2014, Magnussen, Hansen, Planke & Sherson, 2014, Bjælde, Pedersen & Sherson., 2014). Contrary to most crowd science games, Quantum Dreams is not only embedded in the project homepage, but also lives its own life on online app stores with minimal background information. Our "micropayment" is scientific data rather than money. In the game, the players are asked to collect an atom with an optical tweezer and transport it to a target area. A more detailed metagameplay will supporting other play experiences and educational content follow in later iterations.

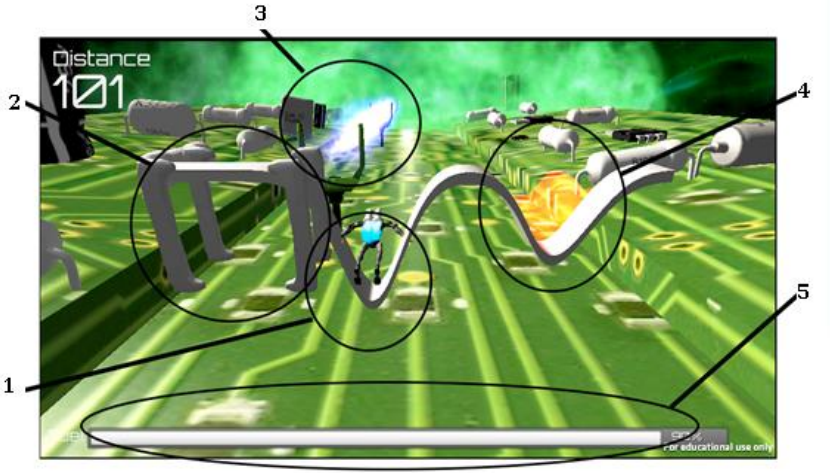

*Figure 1. Quantum Dreams General User Interface (GUI). 1) The optical tweezer which is controlled by the player. The optical tweezer manipulates the atom by changing the potential energy landscape. The robot represents your cursor. 2) The target indicator, which indicates where the target area is going to appear. 3) The target area into which the atom should be moved. When the atom is in the target area, seconds are added to the timer based on the proportion of overlap with the probability distribution. 4) The probability distribution of the atom's location. 5) The timer. When the timer runs out the game is over.*

The GUI resembles Guitar Hero with the player controlling a little flying robot with the mouse flying "into the screen". When a yellow shining substance (figure 1, number 4) appears, the robot can be moved to grab it and ferry it carefully across the screen to hit targets that appear further down the "road" (figure 1, number 3). The yellow substance represents a *probability distribution* of where the atom might be, and the robot controls the optical tweezer. Since atoms in the quantum computer are quite fragile, they must be moved quickly and carefully, or they might be lost due to excitations to high energy states (Sørensen et al., 2015). The game is thus one of fine motor coordination and quickly gleaning the best speed and route, before the robot reaches each target. By repeatedly moving the probability distribution into new target areas during game play, the player helps us map out the best routes in corresponding spaces in the actual quantum computer. The game itself has a technological sciency feel, but the quantum narrative is largely left out of the core loop gameplay itself.

Frames can be understood as the shifting lenses through which we interpret social reality beyond the immediate physical givens (Deterding, 2009; Lieberoth, 2014b). In his seminal work on the subject Erving Goffman (1976) often cites game play as clear example of how people submit to rules and conventions that transform otherwise meaningless actions, such as moving a checkers piece, into significant events within the shared frame of "play". Engrossment into frames oscillates, so as conversation fluxes you might shift attention from meanings within the game, to preserving a friendly relationship with your opponent, and back again (Fine, 1983). Frameworks thus delimit mental and practical situations wherein differing "habits of mind" or "modes of thinking" (Kuhn, 2008) come to the fore. As Quantum Dreams was introduced to our test population in the context of their vocational school, and events started out with a talk on physics, the primary frame of interpretation would have been "education" or "science" for most. The introduction of the highly gamelike GUI, however, *keyed* (as per Goffman, 1976) a swing to "gaming" from which some were not able to shift back. The questions, apart from finding out if the testers enjoyed the game, were thus: Did they remember any physics information? And how do the frames of gaming and science coexist for the players in a simple game experience like this?

**Participants**

38 Danish students (age 14-22, M=17.27, all male) were recruited to play as part of their vocational school (HTX) training. The participants can be described as heavy gamers, with 30 of them reporting playing 10+ hours/week, with high interest in physics (M = 3.74 SD = .852, on a scale 1-5).

**Procedure**

The study took place during an ordinary two-lesson science class at a local vocational school. Participants were informed that they would be part of a usability test for a near-finished crowd science game. The students were first given a presentation of the game, its crowd science purpose, and the underlying physics. The abstract subject matter was adapted to the students' current science-education level. The students then played for 15 minutes on their own laptop computers. After the play session ended, students were given printed surveys as described above. The first page asked them to fill in boxes according to the circles seen in figure 1, describing what each GUI element represented in physics terms. Once done with this task, the students moved on to the likert-style survey.

**Materials**

Participants were given logins to an early version of Quantum Dreams, largely similar to the one launched on Wooglie January 2015. The data were collected with paper surveys. The players were presented with the image of the general user interface (GUI) seen in fig 1, and instructed to *"look at the image. Write in the boxes which physics phenomena the game element represents. If you don't remember the physics term, describe it in your own words. Leave the field empty if you don't remember at all."*

The subsequent pages consisted of a series of multiple-choice questions on a 5-point likert scale from "strongly disagree" to "strongly agree". The scales *interest/enjoyment* (7 items, α = .887), *value/usefulness* (7 items, α = .694), *competence* (6 items, α = .816) and *autonomy* (7 items, α = .760) were adapted from the Intrinsic Motivation Inventories (Intrinsic Motivation Inventory, 1994; Ryan & Deci, 2000). Here, interest/enjoyment is taken to be a main measure of intrinsic motivation stemming from the activity in and of itself, while the other scales are taken to be contributing factors, namely how much the student finds scientific/ educational meaning in the activity, how well they feel that they can do (i.e. mastery) and the degree to which they have flexibility and choice in the participation trajectory. The shorter learning orientation measures in English *mastery* (3 items, α = .285), *performance: approach* (3 items, α = .794), and *performance: avoidance* (3 items, α = .529) were adapted from the Patterns of Adaptive Learning Scales (PALS) (Midgley et al., 2005). These scales are taken to indicate the degree to which learners prefer work that allows for growth through exploration and even constructive failures (mastery) versus just doing well by some objective measure and avoiding looking bad in the eyes of oneself and one's peers (approach/avoid). The scales were supplemented with a series of individual questions mainly used for parts of usability testing that are not reported here. Apart from the PALS-items and the game itself, all questions and instructions were in Danish.

**Data analysis**

Data were analyzed using SPSS 21.0. Central limits theorem assumed for populations over 30. All scales had an acceptable Cronbach's alpha score, except PALS mastery which was abbreviated for an earlier study, and came out with an unacceptable score of .285 (as per Gliem & Gliem, 2003). As a result, it was not used here. Students reported middling performance orientation (M = 3.49, SD = .71) and desire to avoid bad performances (3.12, SD = .69) in their everyday educational lives.

For the GIU-interpretation task all answers were first entered into a spreadsheet, and then, inspired by patterns gleaned by cursory examination of the original paper sheets, a grounded theory approach was used to sort each response into categories according to an open-ended scheme. "Science" and "game" were picked as a priori codes (for a more rigorous example of this technique, see Hoare, Mills, & Francis, 2012). After coding the number of answers attempted, answers in each category, number of correct science answers and number of correct answers in total (even if the task was only to give science answers) were calculated for each participant. A large subset of the students did not attempt to describe any of the GUI-elements, while most of those who did labored to fill in all the boxes. After this exercise, a simple correlation matrix was generated to include the likert items in the analysis.

**Results**

17 out of the 38 students (44.74%) used at least one science explanation to describe a GUI-element. 22 students (57.89%) used at least one game explanation, and 11 students (28.95%) used at least one other kind of conceptualization. The latter conceptual types of answers included descriptions ("*guy who follows the mouse*") or interpretations ("*helper*"). In one instance all of the student's descriptions appeared as unintelligible *1337 speek* gamer slang and abbreviations fit for fast chat channels and message boards. Obviously this kid was deeply engrossed in a gaming mindset, even to a point where he could not (or for identity-reasons opted not to) communicate his interpretations in a way that made sense not just outside the gaming frame, but also outside the culture maintained around hardcore gamer culture. Because no other singular categories emerged in the coding process, descriptive answers that were neither science or game-oriented were grouped together as "conceptual". 11 (28.95%) out of the participant pool left all boxes blank, indicating that they could not find any physics answers as per the instruction, and did not attempt cross-frame explanations in their place. Out of the interpretations given, students on average got two right regardless of category ($M = 2.33$, $SD = 1.27$), but only managed about one correct physics answer ($M = 1.35$, $SD = 1.12$). The number of correct descriptions was obviously dependent on the number of attempts made.

In response to the game experience students' answers indicate above average scores for interest/enjoyment (the main intrinsic motivation measure) (M = 3.67, SD = .55), with slightly lower scores for perceived value/usefulness (M = 3.55, SD = .49), autonomy (M = 3.37, SD = .53) and competence (M = 3.16, S.D. = .62).

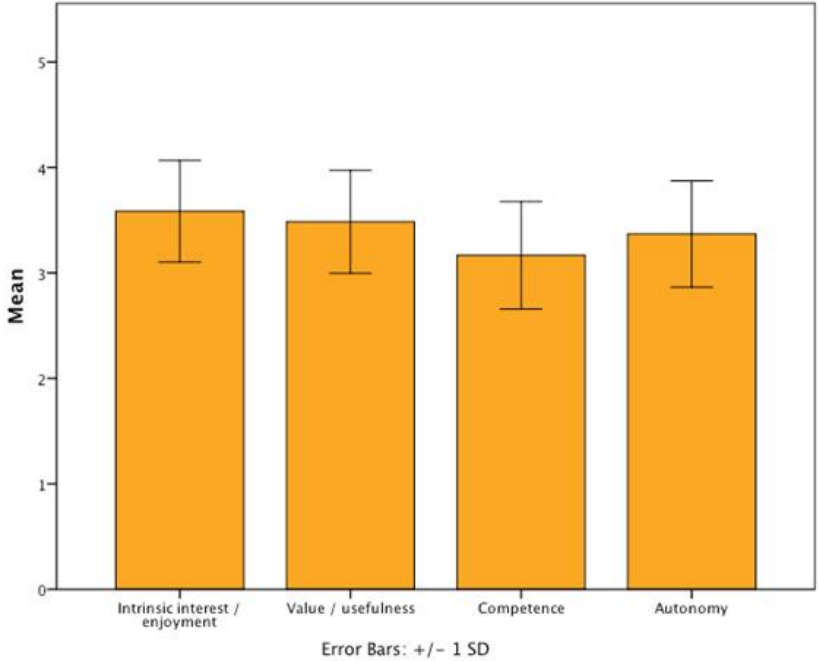

*Figure 2. Intrinsic Motivation Inventory (IMI)*

Quantum Dreams was however *not* perceived as "feeling like other good games" (a validation item used in Lieberoth, 2014b). This was reflected in medium correlations with both interest/enjoyment, $r = .410^{**}$, and value/usefulness, $r = .345^{*}$, and most strongly autonomy $r = .46^{**}$. Physics interest was strongly correlated with interest/enjoyment, $r = .61^{**}$, value/usefulness $r = .533^{**}$, and autonomy, $r = .68^{**}$, as well as a performance approach to learning, $r = .531^{**}$. The PALS scale did not predict any other variables.

When GUI-description categories and precision (i.e. the number of descriptions that could be regarded as accurate) were subsequently also entered into the correlation matrix, autonomy showed up as the only interesting factor: It was very highly correlated with the proportion of correct physics descriptions given, $r = .61**$, while physics interest was only correlated with the general number of correct descriptions given $r = .58**$.

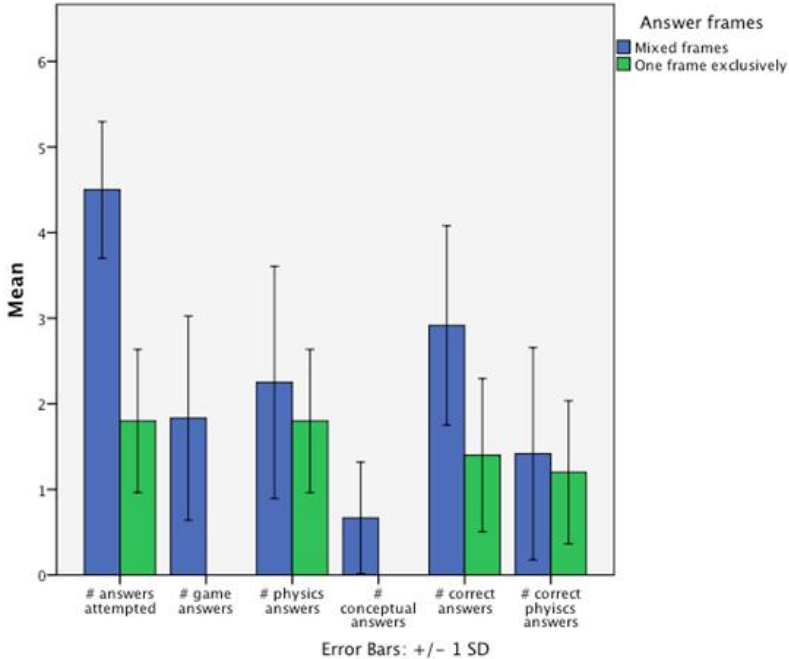

*Figure 3. Answer patterns divided by interpretative frames*

Many used answers from multiple categories to explain GUI-elements, sometimes crossing between them in one answer, but 11 (40.70% of those who attempted any answers) stuck exclusively to one out of the three categories – mostly either game or science. An independent-samples Mann-Whitney test revealed a significant difference between these two groups on number of correct answers** and number of answers attempted**, but not physics answers. Shifting between categories was, however, not negatively related to the total number of correct physics answers achieved either, indicating that flexibly oscillating between frames and thus allowing oneself to give the best answer available at any one point, was an effective way of giving a stream of correct answers overall, without the science understanding suffering – even though the task was to give only science explanations.

**Discussion**

In this essay, we have theorized about the potentials of crowdscience games as opportunities for learning, and described the challenge of several epistemic frames co-existing in the same arena.

In the service of citizen science, a game well played is important on dual dimensions, namely 1.) the purely subjective user experience that, like in any other game, will make people come back for more and tell their friends, and 2.) the quality of data generated results directly engaged players performing at high skill levels. People must literally play the game well, or we will not get the quality of data needed to build our quantum computer.

Game-oriented descriptions were dominant in the vocational class examined here, but this understanding competed with physics thinking. This can be interpreted as a conflict or dynamic oscillation between two prevalent frameworks for interpretation, keyed by elements present in the game experience and the surrounding educational situation.

The importance of the real science subject matter was highlighted by the importance of physics interest and feeling of autonomy. Out of the intrinsic motivation subscales, autonomy stood out as a key variable: It is very possible that we have here gleaned an instance of some students picking between possible frames of engagement, and in the end going directly for the science broccoli.

The dynamics discovered paint an interesting picture of experiences with a game, which can be well played on multiple dimensions – namely both as gaming, learning and participatory science experience. Of course, correlation is not necessarily indicative of learning, neither in the 38-person sample or more generally, and we have no formal before/after tests to show. The game was designed for intrinsically motivated crowdscience participants and not for formal educational deployment, so gains measured at a school like here, would need to be dissociated from the presentation and pedagogy enacted around the play experience. But they paint a strong picture of the mindsets activated around play with a fairly esoteric subject matter, where the main learning must necessarily take place as part of the pedagogies surrounding the experience, even if implicit understandings about the vagaries of quantum particles may be developed through the interactive experience.

It should be noted that our categorization of GUI-element descriptions was based on a rough heuristic categorization. Many of the conceptual descriptions could be argued to have some sort of overlap with the game interpretations, and analyses with more students and questions designed for this end might reveal interesting subcategories. Indeed, the research only came to be written up for publication because interesting patterns emerged from our usability data. We were not aiming to test any particular hypotheses, and did not have clear a priori criteria for data analysis, so the findings here must mainly be viewed as illustrations of relationships between engagement, personal factors (PALS, science interest) and the flux of interpretative frames that guided students' play experience and descriptions of the interface elements.

The patterns seen are encouraging to our claim that crowdscience games hold strong learning potentials, owing to their direct, impactful and *interactive* relationship with *continual* science processes (see also Dewey, 1938b). Indeed, it appears that allowing one's mind to shift between multiple frames of understanding allowed students to come up with descriptions for the physics elements, rather than sticking solely to one mode of explanation and experiencing cognitive roadblocks when the right single-frame answer did not come to mind. But these findings are also a somber reminder that game thinking can be distracting, even when students are explicitly asked to focus on the science explanations. All things considered, many students never supplied any science descriptions, likely owing to the fact that this usability study was not run as part of a continuous educational plan for quantum physics. The pedagogies surrounding any game deployment is likely to be the main contributing factor to student learning, while a game like Quantum Dream supplies a first hand experience with the behavior of atoms in quantum space, which is very hard to grasp even for trained scientists.

**Conclusions**

This was an accidental study. We were looking at user experiences as part of our design process, and found an interesting image of students mixing play and science frames to answer our questions. Some of these discoveries have already been implemented in the game design process, while we are looking deeper into how people cognitively engage with the interface using eye tracking. And of course, the grand prize of implementing play data in quantum physics is an ongoing process.

We have suggested that crowdscience games offer a closer marriage between game and science, but it also looks like these two frames sometimes coexist and sometimes push each other to the side in play trajectories. Our exploration of how students in a vocational class opted to describe different interface elements made the difference between "science", "game" and "conceptual" frameworks of interpretation visible. It appears that the special status of crowdscience games affords some cognitive freedom: An ecology of thinking-layers to oscillate within. This not only supplies multiple routes to engagement but also allows flexible students to exercise a degree of fruitful autonomy in their learning process.

Cooper, S., Khatib, F., Treuille, A., Barbero, J., Lee, J., Beenen, M., … Players, F. (2010). Predicting protein structures with a multiplayer online game. *Nature*, *466*, 756–760. doi:10.1038/nature09304

Davis, P. R., Horn, M. S., & Sherin, B. L. (2013). The Right Kind of Wrong: A "Knowledge in Pieces" Approach to Science Learning in Museums. *Curator: The Museum Journal*, *56*(1), 31–46. doi:10.1111/cura.12005

Deterding, S. (2009). The Game Frame: Systemizing a Goffmanian Approach to Video Game Theory. In *Breaking New Ground: Innovation in Games, Play, Practice and Theory. Proceedings of DiGRA 2009*.

Dewey, J. (1938a). *Experience and education*. New York: Collier.

Dewey, J. (1938b). *Logic: The Theory of Inquiry*. *The Later Works, Volume 12* (pp. 1–18). New York: Hold Rinehart and Winston. doi:10.2307/2180803

Draganov, D. (2014). *Freemium Mobile Games: Design & Monetization* (Kindle edi.). Amazon Digital Services, Inc.

Dreier, O. (2003). Learning in personal trajectories of participation. *Theoretical Psychology: Critical Contributions*, 1–10.

Fields, T. (2014). *Mobile & Social Game Design: Monetization Methods and Mechanics* (2nd ed.). Boca Ranton, FL: CRC press – Taylor & Francis.

Fine, G. A. (1983). *Shared Fantasy: Role Playing Games as Social Worlds*. Chicago: University of Chicago Press.

Gliem, J. a, & Gliem, R. R. (2003). Calculating, Interpreting, and Reporting Cronbach's Alpha Reliability Coefficient for Likert-Type Scales,. *2003 Midwest Research to Practice Conference in Adult,*